\documentstyle{article}
\input{tcilatex}
\QQQ{Language}{American English}

\begin{document}

{\bf \ }

\begin{center}
{\bf MORE ABOUT DONSKER`S DELTA FUNCTION\bigskip\medskip\ \ }

{\bf Angelika Lascheck$^1$}\smallskip\ 

{\bf Peter Leukert$^1$}\smallskip\ 

{\bf Ludwig Streit$^{1,2}$}\smallskip\ 

{\bf Werner Westerkamp$^1$}\medskip\ \ 

$^1$BiBoS - Univ. Bielefeld, D 33615 Bielefeld, Germany\smallskip\ 

{\bf $^2$}Universidade da Madeira, P 9000 Funchal, Portugal
\bigskip \smallskip

Published in Soochow Journal of Mathematics 20 (1994) pp. 401-418
\bigskip \smallskip

{\bf Abstract}:
\end{center}

\begin{quote}
\noindent {\sl We discuss Donsker`s delta function within the framework of
White Noise Analysis, in particular its extension to complex arguments. With
a view towards applications to quantum physics we also study sums and
products of Donsker`s delta functions}.\bigskip\ \smallskip\ 
\end{quote}

\section{{\bf \ }\protect\underline{\QTR{bf}{Introduction}}\ \protect
\medskip
\ \ \ }

White Noise Analysis provides a natural framework for the study of Donsker's
delta function. Thus we review some basic notions and pertinent results from
White Noise Analysis. Then we briefly remark on applications to Feynman
integrals, in particular on the connection between Feynman integrands and
complex scaling.

\noindent Section 3 contains the main results on Donsker's delta. After
extending it to complex parameters we consider its properties under complex
scaling. Then we show how to handle products of delta functions. This has
applications in polymer models, see \cite{Wa1}, \cite{Wa2}, and in series
expansions of Feynman integrands, see \cite{KS} and \cite{LLSW}. Infinite
series of delta functions are also considered. We close this section with a
brief remark on how to define local time within this framework via Donsker's
delta function.

\noindent The final section contains a simple application of some of these
ideas to a quantum mechanical particle on a circle.\bigskip\ \ \ 

\section{\ {\bf \ }\protect\underline{\QTR{bf}{White Noise Analysis}}\protect
\medskip\ \ \ }

{\bf 2.1 Basic notions and results:}\smallskip\ 

\noindent The starting-point of White Noise Analysis is the real Gel`fand
triple\ 

$$
{\cal S}\left( {\bf R}\right) \subset L^2\left( {\bf R}\right) \subset {\cal %
S}^{\prime }\left( {\bf R}\right) , 
$$
where ${\cal S}^{\prime }\left( {\bf R}\right) $ denotes the real Schwartz
space. Using Minlos' theorem we construct the White Noise measure space $%
\left( {\cal S}^{\prime }\left( {\bf R}\right) ,{\cal B},\mu \right) $ by
fixing the characteristic functional in the following way:\ 
$$
C\left( \xi \right) =\int_{{\cal S}^{\prime }\left( {\bf R}\right) }\exp
i\left\langle \omega ,\xi \right\rangle \text{ }d\mu \left( \omega \right)
=\exp \left( -\frac 12\int \xi ^2\left( \tau \right) \;d\tau \right) \text{ }%
,\text{ }\xi \in {\cal S}\left( {\bf R}\right) \ . 
$$
We denote by $\left\langle \cdot ,\cdot \right\rangle $ the bilinear pairing
between ${\cal S}^{\prime }\left( {\bf R}\right) $ and ${\cal S}\left( {\bf R%
}\right) $ and by $\left| \cdot \right| _0$ the norm on $L^2\left( {\bf R}%
\right) $.

\noindent Within this formalism a version of Wiener's Brownian motion is
given by\ 
$$
B\left( t\right) :=\left\langle \omega ,1_{\left[ 0,t\right) }\right\rangle
=\tint\limits_0^t\omega \left( s\right) \text{\thinspace }ds\text{ .} 
$$
\ We now consider the space $\left( L^2\right) $, which is defined to be the
complex Hilbert space \newline $L^2\left( {\cal S}^{\prime }\left( {\bf R}%
\right) ,{\cal B},\mu \right) $. For applications the space $\left(
L^2\right) $ is often too small. A convenient way to solve this problem is
to introduce a space of test functionals in $\left( L^2\right) $ and to use
its larger dual space.

We like to work with the space of test functions $\left( {\cal S}\right) .$
So we review the standard construction of $\left( {\cal S}\right) $ due to 
\cite{KT}. For a more detailed discussion see \cite{HKPS}, \cite{KLPSW}.
Take one system of Hilbertian norms $\left\{ \left| \cdot \right| _p\right\} 
$ topologizing ${\cal S}\left( {\bf R}\right) $ which grows sufficiently
fast. Then ${\cal S}\left( {\bf R}\right) \,$ is realized as a projective
limit of Hilbert spaces ${\cal S}_p\left( {\bf R}\right) :$%
$$
{\cal S}\left( {\bf R}\right) =\bigcap\limits_{p\geq 0}{\cal S}_p\left( {\bf %
R}\right) \ , 
$$
where ${\cal S}_p\left( {\bf R}\right) $ denotes the completition of ${\cal S%
}\left( {\bf R}\right) \,$ w.r.t. $\left| \cdot \right| _p$. Then the space
of tempered distributions is%
$$
{\cal S}^{\prime }\left( {\bf R}\right) =\bigcup\limits_{p\geq 0}{\cal S}%
_{-p}\left( {\bf R}\right) \ , 
$$
where the dual norm $\left| \cdot \right| _{-p}$ topologizes the Hilbert
space ${\cal S}_{-p}\left( {\bf R}\right) $.

One convenient choice is 
\begin{equation}
\label{A}\left| \xi \right| _p:=\left| A^p\xi \right| _0,\quad \xi \in {\cal %
S}\left( {\bf R}\right) , 
\end{equation}
where%
$$
A\xi (t)=-\xi ^{\prime \prime }(t)+\left( t^2+1\right) \xi (t) 
$$
is the Hamiltonian of the harmonic oscillator. Since $\left( L^2\right) $ is
Segal isomorphic to the symmetric Fock space $\Gamma (L^2)$ of $L_{{\bf C}%
}^2\left( {\bf R}\right) :=L^2\left( {\bf R}\right) \oplus iL^2\left( {\bf R}%
\right) $, we can identify the Fock space $\Gamma ({\cal S}_p)$ with a
subspace $({\cal S})_p$ of $\left( L^2\right) $ and define the nuclear space%
$$
\left( {\cal S}\right) =\bigcap_{p\geq 0}\left( {\cal S}\right) _p\ . 
$$
Thus we arrive at the Gel'fand triple:\ 
$$
\left( {\cal S}\right) \subset \left( L^2\right) \subset \left( {\cal S}%
\right) ^{*}. 
$$
Elements of the space $\left( {\cal S}\right) ^{*}$ are called Hida
distributions (or generalized Brownian functionals). It is possible to
characterize the spaces $\left( {\cal S}\right) $ and $\left( {\cal S}%
\right) ^{*}$ by their $S$- or $T$-transforms $\left( \Phi \in \left( {\cal S%
}\right) ^{*},\text{ }\xi \in {\cal S}\left( {\bf R}\right) \right) :$%
\begin{equation}
\label{Tee}T\Phi \left( \xi \right) \equiv \left\langle \!\left\langle \Phi
,\exp \left( i\left\langle \cdot ,\xi \right\rangle \right) \right\rangle
\!\right\rangle =\dint_{{\cal S}^{\prime }\left( {\bf R}\right) }\exp \left(
i\left\langle \omega ,\xi \right\rangle \right) \Phi \left( \omega \right)
d\mu \left( \omega \right) , 
\end{equation}
$$
S\Phi \left( \xi \right) \equiv \left\langle \!\left\langle \Phi ,:\exp
\left\langle \cdot ,\xi \right\rangle :\right\rangle \!\right\rangle , 
$$
here $\left\langle \!\left\langle \cdot ,\cdot \right\rangle \!\right\rangle 
$ denotes the bilinear pairing between $\left( {\cal S}\right) $ and $\left( 
{\cal S}\right) ^{*}$ and we have used the traditional notation\ 

\begin{equation}
\label{Wick}:\exp \left\langle \cdot ,\xi \right\rangle :\text{ }\equiv
C\left( \xi \right) \exp \left( \left\langle \cdot ,\xi \right\rangle
\right) ,\text{ }\xi \in {\cal S}\left( {\bf R}\right) \text{ .} 
\end{equation}
$S$- and $T$-transform have extensions to $\xi \in $ ${\cal S}_{{\bf C}%
}\left( {\bf R}\right) $ and are related by the following formula: 
\begin{equation}
\label{strafo}S\Phi \left( \xi \right) =C\left( \xi \right) \text{ }T\Phi
\left( -i\xi \right) ,\text{ }\xi \in {\cal S}_{{\bf C}}\left( {\bf R}%
\right) \text{ } 
\end{equation}
Let us now quote the above mentioned characterization theorem, which is due
to Potthoff and Streit \cite{PS} and has been generalized in various ways
(see eg. \cite{KLPSW}, \cite{KoS2}, \cite{MY}, \cite{SW}).\bigskip\ \ 
\pagebreak[1]

{\bf Theorem 2.1.1}:\smallskip 

\noindent {\it The following statements are equivalent}:\ \nopagebreak 

\begin{enumerate}
\item  $F:$ ${\cal S}({\bf R)}\rightarrow {\bf C}$ {\it is }\ 

{\bf (A)}{\it \quad {\bf ray-entire}, i.e. for all} $\zeta ,\xi \in {\cal S}%
\left( {\bf R}\right) $ {\it the mapping} $\lambda \mapsto F(\lambda \xi
+\zeta ),\ $\\\TeXButton{10mm}{\hspace*{10mm}}$\lambda \in {\bf R}${\it \
has an entire extension}\ 

{\bf (B)}{\it \quad and uniformly of order two, i.e. there exist constants} $%
K_1,K_2>0$ \\\TeXButton{10mm}{\hspace*{10mm}}{\it such that }%
$$
\left| F\left( z\xi \right) \right| \leq K_1\exp \left( K_2\left| z\right|
^2\left| \xi \right| ^2\right) ,\qquad \forall \xi \in {\cal S}\left( {\bf R}%
\right) ,\ z\in {\bf C}. 
$$
\TeXButton{10mm}{\hspace*{10mm}}{\it for some continuous norm }$\left| \cdot
\right| ${\it \ on }${\cal S}\left( {\bf R}\right) .$

\item  $F$ {\it is the} $S${\it - transform of a Hida distribution} $\Phi
\in \left( {\cal S}\right) ^{*}.$\ 

\item  $F$ {\it is the }$T${\it - transform of a Hida distribution} $%
\stackrel{\wedge }{\Phi }$ $\in \left( {\cal S}\right) ^{*}.$\ 
\end{enumerate}

\noindent Obviously condition (B){\it \ }is implied by condition

{\bf (B')}{\it \quad There exist constants} $K_1,K_2>0$ {\it such that }%
$$
\left| F\left( \xi \right) \right| \leq K_1\exp \left( K_2\left| \xi \right|
^2\right) ,\qquad \forall \xi \in {\cal S}_{{\bf C}}\left( {\bf R}\right) . 
$$
\TeXButton{10mm}{\hspace*{10mm}}{\it for some continuous norm }$\left| \cdot
\right| ${\it \ on }${\cal S}_{{\bf C}}\left( {\bf R}\right) .$

\noindent In the following we will work with condition (B'){\it . }A
functional satisfying 1. is usually called a $U$-functional.\medskip\ 

As an example of an application of this theorem we consider Donsker`s delta
function, the object under consideration in this article.

\noindent Consider the composition $\delta _a\circ B\left( t\right) $ of the
Dirac distribution $\delta _a$ at $a\in {\bf R}$ with Brownian motion $%
B\left( t\right) $, $t>0$:\ 
$$
\Phi =\delta \left( B\left( t\right) -a\right) 
$$
\begin{equation}
\label{Donsker}\Phi =\delta \left( \left\langle \cdot ,1_{\left[ 0,t\right)
}\right\rangle -a\right) ,\text{ }a\in {\bf R}.\text{ } 
\end{equation}
\ The $S$-transform of $\Phi $ is calculated to be \cite{HKPS}, \cite{Ku}:%
$$
S\Phi \left( \xi \right) =\frac 1{\sqrt{2\pi t}}\exp \left( -\frac
1{2t}\left( \dint\limits_0^t\xi \left( s\right) \text{\thinspace }%
ds-a\right) ^{{\bf 2}}\right) 
$$
and theorem 2.1.1 gives immediately that $\Phi $ is a well defined element
in $\left( {\cal S}\right) ^{*}$.\medskip\ \ 

Now we want to mention some important consequences of theorem 2.1.1. The
first one concerns the convergence of sequences of Hida distributions and
can be found in \cite{HKPS}, \cite{PS}$.$\medskip\ 

{\bf Theorem 2.1.2}:\smallskip\ 

\noindent {\it Let} $\left\{ F_n\right\} _{n\in {\bf N}}$ {\it denote a
sequence of }$U${\it -functionals with the following properties:}

\begin{enumerate}
\item  {\it For all} $\xi \in {\cal S}\left( {\bf R}\right) $ , $\left\{
F_n\left( \xi \right) \right\} _{n\in {\bf N}}$ {\it is a Cauchy sequence,}\ 

\item  {\it There exist} $K_{1,}$ $K_2$ $>0$ {\it such that the bound}\ 
$$
\left| F_n\left( z\xi \right) \right| \leq K_1\exp \left( K_2\left| z\right|
^2\left| \xi \right| ^2\right) \text{ },\text{ }\forall \xi \in {\cal S}%
\left( {\bf R}\right) 
$$
{\it holds for almost all }$n\in {\bf N}$ {\it in a continuous norm }$\left|
\cdot \right| ${\it \ on }${\cal S}\left( {\bf R}\right) ${\it . }
\end{enumerate}

\noindent {\it Then there is a unique} $\Phi \in \left( {\cal S}\right) ^{*}$
{\it such that }$T^{-1}F_n$ {\it converges strongly to }$\Phi .$\ \ \ 

\noindent This theorem is also valid for $S$-transforms.\smallskip\ 

Another corollary of theorem 2.1.1 deals with the integration of Hida
distributions which depend on an additional parameter (see \cite{HKPS}, \cite
{KS}).\bigskip
\ \ \ 

{\bf Theorem 2.1.3}:\smallskip\ 

\noindent {\it Let} $\left( \Omega ,B,m\right) $ {\it denote a measure space
and} $\lambda \mapsto \Phi \left( \lambda \right) $ {\it a mapping from} $%
\Omega ${\it \ to} $\left( {\cal S}\right) ^{*}$. {\it Let }$F\left( \lambda
\right) $ {\it denote the} $T$-{\it transform of }$\Phi \left( \lambda
\right) $ {\it which satisfies the following conditions:}\ 

\begin{enumerate}
\item  $\lambda \mapsto F\left( \lambda ,\xi \right) $ {\it is a measurable
function for all} $\xi \in {\cal S}\left( {\bf R}\right) ,$\ 

\item  {\it There exists} {\it a continuous norm }$\left| \cdot \right| $%
{\it \ on }${\cal S}\left( {\bf R}\right) $ {\it such that}%
$$
\ \left| F\text{ }(\lambda ,z\xi )\right| \leq K_1\left( \lambda \right)
\exp \left( K_2\left( \lambda \right) \left| z\right| ^2\left| \xi \right|
^2\right) ,\text{ \quad }\forall \xi \in {\cal S}\left( {\bf R}\right) 
$$
{\it with} $K_1$ $\in $ $L^1\left( \Omega ,m\right) $ {\it and} $K_2\in
L^\infty \left( \Omega ,m\right) .$
\end{enumerate}

\noindent {\it Then} $\Phi $ {\it is Bochner integrable in some }$\left( 
{\cal S}\right) _{-q}$ {\it and thus}\ 
$$
\int\limits_\Omega \Phi \left( \lambda \right) dm\left( \lambda \right) \in
\left( {\cal S}\right) ^{*}. 
$$
{\it Let }$\varphi \in \left( {\cal S}\right) $, {\it then\ }%
$$
\left\langle \!\left\langle \int\limits_\Omega \Phi \left( \lambda \right) 
\text{ }dm\left( \lambda \right) ,\varphi \right\rangle \!\right\rangle
=\int\limits_\Omega \left\langle \!\left\langle \Phi \left( \lambda \right)
,\varphi \right\rangle \!\right\rangle dm\left( \lambda \right) . 
$$
{\it The last equation allows us to intertwine} $T$-{\it transform and
integration}\ 
$$
T\left( \int\limits_\Omega \Phi \left( \lambda \right) \text{ }dm\left(
\lambda \right) \right) \left( \xi \right) =\int\limits_\Omega T\left( \Phi
\left( \lambda \right) \right) \left( \xi \right) \,dm\left( \lambda \right)
. 
$$
\noindent Again the same theorem holds for the $S$-transform.\medskip

Next we shall present a result characterizing $\left( {\cal S}\right) $ in
terms of its $S$-transform (see \cite{kon80}, \cite{KY}, \cite{kps}, \cite
{Lee}, \cite{yan90}).\bigskip\ \ 

\ \ \ {\bf Theorem 2.1.4}:\smallskip\ 

\noindent $F:{\cal S}\left( {\bf R}\right) $ $\rightarrow $ ${\bf C}${\it \
is the }$S${\it -transform of a test functional in} $\left( {\cal S}\right) $%
,{\it \ if and only if }

\begin{description}
\item[(A)]  {\it F is ray-entire,}

\item[(B)]  {\it F is of order 2 and of minimal type, i.e.: For any }$p\in 
{\bf N}$ {\it and} $\forall ${\it \ }$\varepsilon >0${\it ,} {\it there
exists} $K>0${\it \ such that:}%
$$
\left| F(z\xi )\right| \leq K\exp \left( \varepsilon \,\left| z\right|
^2\left| \xi \right| _{-p}^2\right) ,\qquad \forall \xi \in {\cal S}\left( 
{\bf R}\right) ,\ z\in {\bf C}. 
$$
\end{description}

{\bf Remarks}:\\1)\ Theorem 2.1.4 is only valid for the $S$-transform, as a
counter example consider the $T$-transform of $1\in \left( {\cal S}\right) $%
: $T1{}{}\,\left( \xi \right) =\exp \left( -\frac 12\int \xi ^2\left( \tau
\right) \;d\tau \right) ,$ which does not fulfill (B).\\2) In any of the
theorems 2.1.1 to 2.1.4 the growth condition can be replaced by bounds of
the type 2.1.1(B').\bigskip\ \ 

Any element $\varphi \in \left( {\cal S}\right) $ has a pointwise defined
continuous version (see \cite{KY}). Thus it is possible to define a scaling
operator $\sigma _\lambda $ by $\sigma _\lambda \varphi \left( \omega
\right) $ = $\varphi \left( \lambda \omega \right) $\ for $\lambda \in {\bf R%
}$, $\,\omega \in {\cal S}^{\prime }\left( {\bf R}\right) .$ In fact $\sigma
_\lambda $ has an extension to $z\in {\bf C}$ which is a continuous mapping
from $\left( {\cal S}\right) $ into itself (see \cite{HKPS}).\bigskip\ 

{\bf 2.2} {\bf Application to Feynman integrals}\smallskip\\\noindent It has
been shown in \cite{FPS} and \cite{HS} that the kinetic energy term and the
factor compensating the Gaussian fall-off of the White Noise measure combine
to give a well-defined Hida distribution%
$$
I_0=N\!\exp \left( \tfrac{i+1}2\int\nolimits_{{\bf R}}\omega ^2\left( \tau
\right) \,d\tau \right) 
$$
In order to construct the free particle propagator one has to fix the
endpoints of the paths. Within the White Noise framework this can
conveniently be done by multiplying with Donsker`s delta function. This
multiplication of generalized functionals has been justified in \cite{FPS}
and the result $I_0\delta $ has been shown to produce the free particle
propagator. A more elegant and general way of defining products of $I_0$ and
other distributions has been suggested in \cite{S}, where the connection
between $I_0$ and complex scaling was noted. To see this consider the $T$%
-transform of $I_0$: 
$$
TI_0\left( \xi \right) =\exp \left( -\tfrac i2\tint\nolimits_{{\bf R}}\xi
^2\left( \tau \right) \,d\tau \right) ={\bf E}\left( \exp \left(
i\left\langle \omega ,\sqrt{i}\xi \right\rangle \right) \right) 
$$
$$
={\bf E}\left( \sigma _{\sqrt{i}}^{\dagger }1\cdot e^{i\left\langle \omega
,\xi \right\rangle }\right) =\left( T\sigma _{\sqrt{i}}^{\dagger }1\right)
\left( \xi \right) \text{.} 
$$
Thus one has $I_0=\sigma _{\sqrt{i}}^{\dagger }1$ by theorem 2.1.1.

\noindent In order to define products by $I_0$ one approximates the other
factor by test functionals and then studies the convergence of the scaled
sequence according to the following theorem from \cite{S} (see also \cite
{HKPS}):\medskip\ 

{\bf Theorem 2.2.1.}:\smallskip\ 

\noindent {\it Let }$\varphi _n$ {\it be a sequence of test functionals.
Then the following statements are equivalent:}\smallskip\ \ 

$\left( i\right) $ {\it The sequence} $I_0\varphi _n\rightarrow \Psi $ {\it %
converges in} $\left( {\cal S}\right) ^{*}$.\smallskip\ \ \ 

$\left( ii\right) $ {\it The sequence} $\sigma _{\sqrt{i}}\varphi _n$ {\it %
converges in} $\left( {\cal S}\right) ^{*}$.\smallskip\ \ \ 

$\left( iii\right) $ {\it The sequence} ${\bf E}\left( \psi \,\sigma _{\sqrt{%
i}}\varphi _n\right) $ {\it converges for all} $\psi \in \left( {\cal S}%
\right) $.\smallskip\ \ \ 

\noindent {\it The action of} $\Psi $ {\it is given by }$\left\langle
\!\left\langle \Psi ,\psi \right\rangle \!\right\rangle =\stackunder{%
n\rightarrow \infty }{\lim }{\bf E}\left( \sigma _{\sqrt{i}}\left( \varphi
_n\psi \right) \right) $,

\noindent {\it if one of the conditions }$\left( i\right) ${\it \ to} $%
\left( iii\right) ${\it \ holds}.\bigskip\ \ \ 

\section{{\bf \ }\protect\underline{\QTR{bf}{Donsker`s Delta Function}}%
\protect\medskip\ \ }

\ \noindent Consider again the $S$-transform of Donsker's delta function:%
$$
S\Phi \left( \xi \right) =\frac 1{\sqrt{2\pi t}}\exp \left( -\frac
1{2t}\left( \dint\limits_0^t\xi \left( s\right) \text{\thinspace }%
ds-a\right) ^{{\bf 2}}\right) \text{ .} 
$$
This is clearly analytic in the parameter $a\in {\bf R}$. We can thus extend
to complex $a$ and the resulting expression is still a $U$-functional. Hence
by theorem 2.1.1 it is possible to define $\delta \left( B(t)-a\right) $ for 
$a\in {\bf C}$.\bigskip\ \ 

{\bf 3.1} {\bf Complex Scaling}\smallskip\ \ \ 

\noindent We intend to study complex scaling of a sequence of test
functionals converging to $\delta .$ Let $\eta _n\in {\cal S}\left( {\bf R}%
\right) $ be a sequence of real Schwartz test functions converging to $\eta
\equiv 1_{\left[ 0,t\right) }$ in $L^2\left( {\bf R}\right) .$

\noindent Choose $\left| \alpha \right| <\frac \pi 4$ and $z\in {\bf S}%
_\alpha \equiv \left\{ z\in {\bf C}\mid \arg z\in \left( -\frac \pi 4+\alpha
,\frac \pi 4+\alpha \right) \right\} $ and define 
\begin{equation}
\label{fin}\varphi _{n,z}\left( \omega \right) =\frac 1{2\pi
}\int\limits_{-ne^{-i\alpha }}^{ne^{-i\alpha }}e^{i\lambda \left(
z\left\langle \omega ,\eta _n\right\rangle -a\right) }d\lambda \text{ .} 
\end{equation}
To shorten notation we call the basic sequence $\varphi _{n,1}$ simply $%
\varphi _n$. Note that given any $z$ with $\left| \arg z\right| <\frac \pi 2$
one can choose $\alpha $ such that $z$ and $1$ are in ${\bf S}_\alpha $. In
this section we will establish the following results:\medskip 
\ 

{\bf Theorem 3.1.1:}\smallskip 

\noindent {\it For all} $z\in {\bf S}_\alpha $ {\it we have:\ }

\begin{description}
\item[(i)]  $\varphi _{n,z}\in \left( {\cal S}\right) $.

\item[(ii)]  $\sigma _z\varphi _n=\varphi _{n,z}$.

\item[(iii)]  $\varphi _n\rightarrow \delta $ {\it in }$\left( {\cal S}%
\right) ^{*}$.

\item[(iv)]  $\sigma _z\varphi _n$ {\it converges in} $\left( {\cal S}%
\right) ^{*}$.\\\TeXButton{10mm}{\hspace*{10mm}}{\it The limit element is
called }$\sigma _z\delta $.

\item[(v)]  $\sigma _z\delta $ $\in \left( {\cal S}\right) _{-q}$ {\it for }$%
q>\frac 12,$\\\TeXButton{10mm}{\hspace*{10mm}}{\it with the choice of of
norms according to (\ref{A}).}

\item[(vi)]  $\delta $ {\it is homogenuous of degree} $-1$ : \\%
\TeXButton{10mm}{\hspace*{10mm}}$\sigma _z\delta \left( \left\langle \omega
,\eta \right\rangle -a\right) =\frac 1z\delta \left( \left\langle \omega
,\eta \right\rangle -\frac az\right) .$
\end{description}

\noindent {\bf Remark:} The limit elements in (iii) and (iv) do not depend
on $\alpha .$\medskip\ 

{\bf Proposition 3.1.2}:\smallskip\ 

\noindent {\it For all} $z\in {\bf S}_\alpha $ {\it we have} 
$$
\varphi _{n,z}\left( \omega \right) \in \left( {\cal S}\right) \text{ .} 
$$

\ {\bf Proof}:

\noindent First of all we calculate the $S$-transform of the integrand of
equation (\ref{fin}):%
$$
S\left( \exp \left( i\lambda \left( z\left\langle \omega ,\eta
_n\right\rangle -a\right) \right) \right) \left( \xi \right) 
$$
$$
=\exp \left( -\frac 12z^2\lambda ^2\left| \eta _n\right| _0^2+i\lambda
\left( z\left( \xi ,\eta _n\right) -a\right) \right) \text{ .} 
$$
To apply theorem 2.1.3 we have to find an estimate for the $S$-transform%
$$
\left| S\left( \exp \left( i\lambda \left( z\left\langle \omega ,\eta
_n\right\rangle -a\right) \right) \right) \left( \xi \right) \right| 
$$
$$
\leq \exp \left( \frac 12\left| z^2\right| \left| \lambda ^2\right| \left|
\eta _n\right| _0^2+\left| \lambda \right| \left| z\right| \left| \xi
\right| _0\left| \eta _n\right| _0+\left| \lambda \right| \left| a\right|
\right) 
$$
$$
\leq \exp \left( \left| \lambda ^2\right| \left| z\right| ^2\left| \eta
_n\right| _0^2+\left| \lambda \right| \left| a\right| \right) \exp \left(
\frac 12\left| \xi \right| _0^2\right) \text{ }\forall \xi \in {\cal S}_{%
{\bf C}}({\bf R)} 
$$
This fulfills the requirements of theorem 2.1.3, thus the integral (\ref{fin}%
) is well-defined.%
\begin{eqnarray*}
S\varphi _{n,z}\left( \xi \right) &=&\frac 1{2\pi }\int\limits_{-ne^{-i\alpha }}^{ne^{-i\alpha }}S\left( \exp \left( i\lambda \left( z\left\langle \omega ,\eta _n\right\rangle -a\right) \right) \right) \left( \xi \right) d\lambda \\
&=&\frac 1{2\pi }\int\limits_{-ne^{-i\alpha }}^{ne^{-i\alpha }}\exp \left( -\frac 12z^2\lambda ^2\left| \eta _n\right| _0^2+i\lambda \left( z\left( \xi ,\eta _n\right) -a\right) \right) d\lambda \text{ .}
\end{eqnarray*} We substitute $\nu =e^{i\alpha }\lambda $ , this leads to\ \ 
\begin{equation}
\label{esfi}S\varphi _{n,z}\left( \xi \right) =\frac 1{2\pi
}\int\limits_{-n}^n\exp \left( -\frac 12z^2e^{-2i\alpha }\nu ^2\left| \eta
_n\right| _0^2+ie^{-i\alpha }\nu \left( z\left( \xi ,\eta _n\right)
-a\right) \right) e^{-i\alpha }d\nu . 
\end{equation}
Now take the absolut value%
\begin{eqnarray*}
  \left| S\varphi _{n,z}\left( \xi \right) \right|  & \leq &\frac 1{2\pi }\int\limits_{-n}^n
\exp \left( \frac 12\left| z\right| ^2\nu ^2\left| \eta _n\right| _0^2
+\left| \nu \right| \left| z\right| \left| \xi \right|_{-p} \left|  \eta _n \right|_p 
+\left| \nu \right| \left| a\right| \right) d\nu \\
  & \leq  & \frac 1{2\pi }\int\limits_{-n}^n\exp \left( \frac 12\left| z\right| ^2\nu ^2
\left| \eta _n\right| _0^2+\frac 1{2s^2}\nu ^2\left| z\right| ^2\left| \eta _n\right|_p ^2
+\frac 12s^2\left| \xi \right|_{-p}^{2}+\left| \nu \right| \left| a\right| \right) d\nu \\
  & \leq  & \frac n\pi \exp \left( \frac{n^2}2\left| z\right| ^2\left( 
\left| \eta_n \right|_0^2+\frac 1{s^2} \left| \eta_n \right|_p^2\right) 
+n\left| a\right| \right) \exp\left(+\frac 12s^2\left| \xi \right|_{-p}^{2}\right)\text{ .}
\end{eqnarray*}This estimate holds for all $s\in {\bf R}$ and $p\in {\bf N}.$
Thus it fulfills the requirements of the characterization theorem 2.1.4 and
we arrive at $\varphi _{n,z}\in \left( {\cal S}\right) .$%
\TeXButton{End Proof}{\endproof}\medskip\ \ 

\noindent Now we study the action of $\sigma _z$ on $\varphi _n$. This leads
to the following\bigskip\ 

{\bf Proposition 3.1.3}:%
$$
\sigma _z\varphi _n\left( \omega \right) =\varphi _{n,z}\left( \omega
\right) \text{ , }z\in {\bf S}_\alpha \text{.} 
$$

{\bf Proof}:

\noindent A direct computation yields%
$$
\varphi _n\left( \omega \right) =\frac 1{2\pi }\int\limits_{-ne^{-i\alpha
}}^{ne^{-i\alpha }}e^{i\lambda \left( \left\langle \omega ,\eta
_n\right\rangle -a\right) }d\lambda 
$$
$$
=\frac 1{2\pi i\left( \left\langle \omega ,\eta _n\right\rangle -a\right)
}\left( \exp \left( ine^{-i\alpha }\left( \left\langle \omega ,\eta
_n\right\rangle -a\right) \right) -\exp \left( -ine^{-i\alpha }\left(
\left\langle \omega ,\eta _n\right\rangle -a\right) \right) \right) 
$$
$$
=\frac 1{\pi \left( \left\langle \omega ,\eta _n\right\rangle -a\right)
}\sin \left( ne^{-i\alpha }\left( \left\langle \omega ,\eta _n\right\rangle
-a\right) \right) \text{ .} 
$$
This is defined pointwise and continuous. Thus%
$$
\sigma _z\varphi _n\left( \omega \right) =\frac 1{\pi \left( z\left\langle
\omega ,\eta _n\right\rangle -a\right) }\sin \left( ne^{-i\alpha }\left(
z\left\langle \omega ,\eta _n\right\rangle -a\right) \right) \text{ .} 
$$
On the other hand we get%
$$
\frac 1{2\pi }\int\limits_{-ne^{-i\alpha }}^{ne^{-i\alpha }}e^{i\lambda
\left( z\left\langle \omega ,\eta _n\right\rangle -a\right) }d\lambda \text{ 
}=\frac 1{\pi \left( z\left\langle \omega ,\eta _n\right\rangle -a\right)
}\sin \left( ne^{-i\alpha }\left( z\left\langle \omega ,\eta _n\right\rangle
-a\right) \right) =\varphi _{n,z}\left( \omega \right) \text{%
\TeXButton{nopagebreak}{\nopagebreak}.} 
$$
\TeXButton{End Proof}{\endproof}

\ 

{\bf Proposition 3.1.4}:\medskip

\noindent {\it Let }$z\in {\bf S}_\alpha $.

\begin{description}
\item[(i)]  $\sigma _z\varphi _n\in \left( {\cal S}\right) $ $\forall n\in 
{\bf N}$.

\item[(ii)]  $\varphi _n\rightarrow \delta $ {\it in }$\left( {\cal S}%
\right) ^{*}$.

\item[(iii)]  $\sigma _z\varphi _n$ {\it converges in} $\left( {\cal S}%
\right) ^{*}$.\\\noindent  {\it The limit element of} $\sigma _z\varphi _n$ 
{\it is called} $\sigma _z\delta $.\medskip\ 
\end{description}

{\bf Proof}:\smallskip\ 

\noindent The first statement is clear from proposition 3.1.2 and
proposition 3.1.3.

\noindent For the proof of (iii) let us look at (\ref{esfi}) again. In order
to use the convergence theorem, we have to find a bound of (\ref{esfi}):%
$$
\left| S\sigma _z\varphi _n\left( \xi \right) \right| 
$$
$$
\leq \frac 1{2\pi }\int\limits_{-\infty }^\infty \exp \left( -\frac
12Re\left( z^2e^{-2i\alpha }\right) \nu ^2\left| \eta _n\right| _0^2+\nu
Re\left( ie^{-i\alpha }\left( z\left( \xi ,\eta _n\right) -a\right) \right)
\right) d\nu \text{ .} 
$$
The integral exists if $Re\left( z^2e^{-2i\alpha }\right) >0$.

\noindent This condition is satisfied for $-\frac \pi 4+\alpha <\arg z<\frac
\pi 4+\alpha $ . We get%
$$
\left| S\sigma _z\varphi _n\left( \xi \right) \right| \leq \frac 1{2\pi }%
\sqrt{\frac{2\pi }{Re\left( z^2e^{-2i\alpha }\right) \left| \eta _n\right|
_0^2}}\exp \left( \frac{\left[ Re\left( ie^{-i\alpha }\left( z\left( \xi
,\eta _n\right) -a\right) \right) \right] ^2}{2Re\left( z^2e^{-2i\alpha
}\right) \left| \eta _n\right| _0^2}\right) \text{ } 
$$
$$
\leq \frac 1{\sqrt{2\pi Re\left( z^2e^{-2i\alpha }\right) \frac 12\left|
\eta \right| _0^2}}\exp \left( \frac{\left( \left| z\right| \left| \xi
\right| _02\left| \eta \right| _0+\left| a\right| \right) ^2}{2Re\left(
z^2e^{-2i\alpha }\right) \frac 12\left| \eta \right| _0^2}\right) \text{ ,} 
$$
for $n$ large enough.

\noindent Now the convergence theorem applies:%
\begin{eqnarray*}
\lim \limits_{n\rightarrow \infty }S\sigma _z\varphi _n\left( \xi \right) &=&\frac 1{2\pi }e^{-i\alpha }\int\limits_{-\infty }^\infty \exp \left( -\frac 12z^2e^{-2i\alpha }\nu ^2\left| \eta \right| _0^2+i\nu e^{-i\alpha }\left( z\left( \xi ,\eta \right) -a\right) \right) d\nu \\
&=&e^{-i\alpha }\frac 1{\sqrt{2\pi }ze^{-i\alpha }\left| \eta \right| _0}\exp \left( \frac{-e^{-2i\alpha }\left( z\left( \xi ,\eta \right) -a\right) }{2\left| \eta \right| _0^2z^2e^{-2i\alpha }}^2\right) \\
&=&\frac 1{\sqrt{2\pi }z\left| \eta \right| _0}\exp \left( \frac{-\left( z\left( \xi ,\eta \right) -a\right) }{2\left| \eta \right| _0^2z^2}^2\right) \text{ .}
\end{eqnarray*}Note that the limit does not depend on $\alpha $.

\noindent To prove (ii) we simply set $z=1$ in the above formula.%
\TeXButton{End Proof}{\endproof}

\ 

{\bf Proposition 3.1.5}:\smallskip\ 

\noindent $\delta $ {\it is homogenuous of degree} $-1${\it \ in} $z\in {\bf %
S}_\alpha $:%
$$
\sigma _z\delta \left( \left\langle \omega ,\eta \right\rangle -a\right)
=\frac 1z\delta \left( \left\langle \omega ,\eta \right\rangle -\frac
az\right) . 
$$

{\bf Proof}:\\\noindent From the last formula in the proof of 3.1.4 we have: 
\begin{equation}
\label{sz}S\sigma _z\delta \left( \left\langle \omega ,\eta \right\rangle
-a\right) \left( \xi \right) =\frac 1{\sqrt{2\pi }z\left| \eta \right|
_0}\exp \left( -\frac 12\frac{\left( a-z\left( \xi ,\eta \right) \right) ^2}{%
z^2\left| \eta \right| _0^2}\right) 
\end{equation}
$$
=\frac 1z\frac 1{\sqrt{2\pi }\left| \eta \right| _0}\exp \left( -\frac 12%
\frac{\left( \frac az-\left( \xi ,\eta \right) \right) ^2}{\left| \eta
\right| _0^2}\right) =S\left( \frac 1z\delta \left( \left\langle \omega
,\eta \right\rangle -\frac az\right) \right) \left( \xi \right) \text{ .%
\TeXButton{End Proof}{\endproof}\ } 
$$

The following proposition gives the degree of singularity of the scaled
Donsker's delta.\medskip\ 

{\bf Proposition 3.1.6}:\smallskip\ 

\noindent $\sigma _z\delta $ $\in \left( {\cal S}\right) _{-q}$ {\it for} $%
q>\frac 12$, $z\in {\bf S}_\alpha ${\it , with the choice of norms} {\it (%
\ref{A}).}\medskip\ \ \ 

{\bf Proof}:\ \\\noindent We use the $S$-transform to get the following
estimate:%
$$
\left| S\left( \sigma _z\delta \right) \left( \xi \right) \right| =\frac 1{%
\sqrt{2\pi }\left| z\right| \left| \eta \right| _0}\left| \exp \left( \frac{%
-\left( a-z\left( \xi ,\eta \right) \right) }{2\left| \eta \right| _0^2z^2}%
^2\right) \right| 
$$
\ 
$$
\leq \frac 1{\sqrt{2\pi }\left| z\right| \left| \eta \right| _0}\exp \left(
\frac 12\left| \xi \right| _0^2+\frac{\left| a\right| }{\left| \eta \right|
_0\left| z\right| }\left| \xi \right| _0+\frac 12\frac{\left| a\right| ^2}{%
\left| \eta \right| _0^2\left| z\right| ^2}\right) 
$$
$$
\leq \frac 1{\sqrt{2\pi }\left| z\right| \left| \eta \right| _0}\exp \left( 
\frac{\left( 1+\frac 1{s^2}\right) \left| a\right| ^2}{2\left| \eta \right|
_0^2\left| z\right| ^2}\right) \exp \left( \frac 12\left( 1+s^2\right)
\left| \xi \right| _0^2\right) \text{ \quad }\forall \text{ }s>0. 
$$
An estimate in \cite{PS2} shows, that $\Phi \in \left( {\cal S}\right) _{-q}$
for any $q>\frac{\ln K_2}{2\ln 2}+p+1$ if\\ $\left| F_n\left( z\xi \right)
\right| \leq K_1\exp \left( K_2\left| z\right| ^2\left| \xi \right|
_p^2\right) $ $,$ $\forall \xi \in {\cal S}\left( {\bf R}\right) $ . Thus $%
\sigma _z\delta \in \left( {\cal S}\right) _{-q}$ for $q>\frac{\ln \frac
12\left( 1+s^2\right) }{2\ln 2}+1.$ As $s$ can be chosen arbitrarily small,
we may use any $q>\frac 12$.\TeXButton{End Proof}{\endproof}

\ \ 

{\bf Corollary 3.1.7}:\smallskip\ 

\noindent {\it The mapping} $z\mapsto $$\sigma _z\delta $ {\it is continuous
from} ${\bf S}_\alpha $ {\it to} $\left( {\cal S}\right) _{-q}$ $,q>\frac 12$%
.\medskip\ \ 

{\bf Proof}:\smallskip\ 

\noindent Choose a sequence $z_n\rightarrow z$ in ${\bf S}_\alpha .$ We have
the bound%
$$
\left| S\sigma _{z_n}\delta \left( \xi \right) \right| \leq \frac 1{\sqrt{%
2\pi }\left| z_n\right| \left| \eta \right| _0}\exp \left( \frac{\left(
1+\frac 1{s^2}\right) \left| a\right| ^2}{2\left| \eta \right| _0^2\left|
z_n\right| ^2}\right) \exp \left( \frac 12\left( 1+s^2\right) \left| \xi
\right| _0^2\right) . 
$$
Define $\frac 1{\sqrt{2\pi }\left| z_n\right| \left| \eta \right| _0}\exp
\left( \frac{\left( 1+\frac 1{s^2}\right) \left| a\right| ^2}{2\left| \eta
\right| _0^2\left| z_n\right| ^2}\right) \equiv K_{1,n}$, from the
convergence of the $K_{1,n}$ the existence of a uniform bound for $\left|
S\sigma _{z_n}\delta \left( \xi \right) \right| $ can be deduced. Now
theorem 2.1.2 ensures convergence of the sequence $\left\{ \sigma
_{z_n}\delta \right\} $ in $\left( {\cal S}\right) _{-q}$.%
\TeXButton{End Proof}{\endproof}\bigskip
\medskip\ 

{\bf 3.2} {\bf Products of Donsker`s Delta Functions}\bigskip\ \\\noindent %
To define products of scaled Donsker`s deltas, we use the following ansatz 
$$
\Phi =\tprod\limits_{j=1}^n\sigma _z\delta \left( \left\langle \cdot
,f_j\right\rangle -a_j\right) =\frac 1{\left( 2\pi \right)
^n}\tprod\limits_{j=1}^n\dint\limits_\gamma \exp \left( i\lambda _j\left(
z\left\langle \cdot ,f_j\right\rangle -a_j\right) \right) d\lambda _j\ , 
$$
here $\gamma =\left\{ e^{-i\alpha }t\mid t\in {\bf R}\right\} $, $z\in {\bf S%
}_\alpha $, where $\alpha $ is choosen such that $\left| \alpha \right|
<\frac \pi 4$, $f_j$ are real, linear independent elements of $L^2$ and $%
a_j\in {\bf C}$. We use the notation

$\exp \left( \sum\limits_{j=1}^ni\lambda _j\left( \left\langle \cdot
,f_j\right\rangle -a_j\right) \right) =\exp \left( i\overrightarrow{\lambda }%
\left( \left\langle \cdot ,\overrightarrow{f}\right\rangle -\overrightarrow{a%
}\right) \right) $.

\noindent To prove that $\Phi $ is well-defined, we calculate it`s $T$%
-transform:%
\begin{eqnarray*}
\ T\Phi \left( \xi \right)  & = & \frac{e^{-i\alpha n}}{\left( 2\pi \right) ^n}
\dint \dint \exp \left( ize^{-i\alpha }\left\langle \omega ,\overrightarrow{\lambda }
\overrightarrow{f}\right\rangle -ie^{-i\alpha }\overrightarrow{\lambda }
\overrightarrow{a} +i\left\langle \omega ,\xi \right\rangle \right) d\mu 
\,d^n\lambda  \\
& = & \frac 1{\left( 2\pi \right) ^n}e^{-i\alpha n}\dint \dint \exp \left( i\left\langle
 \omega ,ze^{-i\alpha }\overrightarrow{\lambda }\overrightarrow{f}+\xi \right
\rangle -ie^{-i\alpha }\overrightarrow{\lambda }\overrightarrow{a}\right) d\mu 
\, d^n\lambda  \\
& = & \frac 1{\left( 2\pi \right) ^n}e^{-i\alpha n}\dint C\left( ze^{-i\alpha }
\overrightarrow{\lambda }\overrightarrow{f}+\xi \right) \exp \left(-ie^{-i\alpha }
\overrightarrow{\lambda }\overrightarrow{a} \right) d^n\lambda  \\
&=&\frac 1{\left( 2\pi \right) ^n}e^{-i\alpha n}\dint \exp \left[ -\frac 12\dint 
\left( ze^{-i\alpha }\overrightarrow{\lambda }\overrightarrow{f}+\xi \right) ^2
d\tau -ie^{-i\alpha }\overrightarrow{\lambda }\overrightarrow{a}\right] 
d^n\lambda \text{ .}
\end{eqnarray*}
\\Consider now%
$$
\dint \left( ze^{-i\alpha }\overrightarrow{\lambda }\overrightarrow{f}(\tau
)\right) ^2d\tau =z^2e^{-i2\alpha }\dsum\limits_{k,l}\lambda _k\lambda
_l\left( f_k,f_l\right) =z^2e^{-i2\alpha }\overrightarrow{\lambda }M%
\overrightarrow{\lambda }, 
$$
where $M\equiv \left( f_k,f_l\right) _{k,l}$. This is a Gram matrix of
linear independent vectors and thus positive definite.%
\begin{eqnarray*}
  T\Phi \left( \xi \right) \;& = &\frac 1{\left( 2\pi \right) ^n}\,e^{-i\alpha n}
e^{-\frac 12\int \xi ^2\left( \tau \right) \;d\tau } \\
  & &\times \dint \exp \left[ -\frac 12z^2e^{-i2\alpha }\overrightarrow{\lambda }
M\overrightarrow{\lambda }-ze^{-i\alpha }\overrightarrow{\lambda }
\left( \overrightarrow{f},\xi \right) -ie^{-i\alpha }\overrightarrow{\lambda }
\overrightarrow{a}\right] d^n\lambda \\
  & = &\sqrt{\frac{\left( 2\pi \right) ^n}{\left( z^2e^{-i2\alpha }\right) ^n\det M}}\,
\frac{e^{-i\alpha n}}{\left( 2\pi \right) ^n}\;e^{-\frac 12\int \xi ^2\left( \tau \right)
 \;d\tau }
\end{eqnarray*}
$$
\times \exp \left[ \frac 12\left( ze^{-i\alpha }\left( \overrightarrow{f}%
,\xi \right) +ie^{-i\alpha }\overrightarrow{a}\right) \left( z^2e^{-i2\alpha
}M\right) ^{-1}\left( ze^{-i\alpha }\left( \overrightarrow{f},\xi \right)
+ie^{-i\alpha }\overrightarrow{a}\right) \right] , 
$$
\ \ this Gaussian integral exists if $Re\left( z^2e^{-i2\alpha }\right) >0$,
which is equivalent to $z\in {\bf S}_\alpha $. The last expression is a $U$%
-functional, so we get:

\ 

{\bf Theorem 3.2.1}:\smallskip\ 

\noindent {\it Let} $a_j\in {\bf C}$, $f_j\in L^2({\bf R)}$ {\it linear
independent and} $M$ $=\left( f_k,f_l\right) _{k,l}$ {\it the corresponding
Gram matrix.}

\noindent {\it Then for all }$z\in {\bf S}_\alpha $ $\Phi
=\tprod\limits_{j=1}^n\sigma _z\delta \left( \left\langle \cdot
,f_j\right\rangle -a_j\right) $ {\it is a Hida distribution with} $S${\it %
-transform}%
$$
S\Phi \left( \xi \right) =\frac 1{\sqrt{\left( 2\pi z^2\right) ^n\det M}%
}\exp \left[ -\frac 12\left( \left( \overrightarrow{f},\xi \right) -\frac 1z%
\overrightarrow{a}\right) M^{-1}\left( \left( \overrightarrow{f},\xi \right)
-\frac 1z\overrightarrow{a}\right) \right] \text{ .\bigskip}\ 
$$

\bigskip\ {\bf 3.3 Series of Donsker`s Delta Functions}:\smallskip\ \\%
\noindent We set 
$$
\Phi _N=\sum\limits_{n=-N}^N\sigma _z\,\delta \left( B\left( t\right)
-a+n\right) ,\ \text{ }a\in {\bf C}\text{ .} 
$$
This is a well-defined Hida distribution and its $S$-transform is given by 
$$
S\Phi _N\left( \xi \right) =\frac 1{\sqrt{2\pi t}z}\sum\limits_{n=-N}^N\exp
\left( -\frac 1{2tz^2}\left( a-n-z\int\limits_0^t\xi \left( s\right)
\,ds\right) ^2\right) \text{ .} 
$$
We now assume $Re\frac 1{z^2}>0.$ To study the limit $N\rightarrow \infty $
we try to find a uniform bound (in $N$) for 
$$
\left| S\Phi _N\left( \xi \right) \right| \leq \frac 1{\left| z\right| \sqrt{%
2\pi t}}\sum\limits_{n=-N}^N\exp \left( -\frac 1{2t}Re\left( \frac
1{z^2}\left( a-n-z\int\limits_0^t\xi \left( s\right) \,ds\right) ^2\right)
\right) \text{ } 
$$
$$
\leq \frac 1{\left| z\right| \sqrt{2\pi t}}\sum\limits_{n=-N}^N\exp \left(
\frac 1{2t}\left( -n^2Re\frac 1{z^2}+2\left| \frac nz\right| \left| \frac
az-\int\limits_0^t\xi \left( s\right) \,ds\right| +\left| \frac
az-\int\limits_0^t\xi \left( s\right) \,ds\right| ^2\right) \right) 
$$
$$
\leq \frac 1{\left| z\right| \sqrt{2\pi t}}\sum\limits_{n=-\infty }^\infty
\exp \left( \frac 1{2t}\left( -n^2\frac 12Re\frac 1{z^2}+\left( 1+\frac{%
2\left| z\right| ^2}{Re\ z^2}\right) \left| \frac az-\int\limits_0^t\xi
\left( s\right) \,ds\right| ^2\right) \right) 
$$
$$
=\frac 1{\left| z\right| \sqrt{2\pi t}}\exp \left( \frac 1{2t}\left( 1+\frac{%
2\left| z\right| ^2}{Re\ z^2}\right) \left( \left| \frac az\right| -\sqrt{t}%
\left| \xi \right| _0\right) ^2\right) \sum\limits_{n=-\infty }^\infty \exp
\left( -\frac 1{4t}Re\frac 1{z^2}n^2\right) \text{.} 
$$
The infinite sum converges if $Re\frac 1{z^2}>0$, i.e. if $z\in {\bf S}_0$.
The sum can also be expressed using the theta function (see \cite{Mu}) $%
\vartheta \left( \rho ,\tau \right) =\sum\limits_{n=-\infty }^\infty \exp
\left( \pi in^2\tau +2\pi in\rho \right) $ as $\vartheta \left( 0,\frac
i{4\pi t}Re\frac 1{z^2}\right) $.

\noindent Since now the convergence theorem applies, we get:%
$$
S\Phi \left( \xi \right) =\frac 1{z\sqrt{2\pi t}}\sum\limits_{n=-\infty
}^\infty \exp \left( -\frac 1{2tz^2}\left( a-n-z\int\limits_0^t\xi \left(
s\right) \,ds\right) ^2\right) 
$$
$$
=\frac 1{z\sqrt{2\pi t}}\exp \left( -\frac 1{2t}\left( \int\limits_0^t\xi
\left( s\right) \,ds-\frac az\right) ^2\right) \sum\limits_{n=-\infty
}^\infty \exp \left( -\frac{n^2}{2tz^2}-\frac nt\left( \int\limits_0^t\xi
\left( s\right) \,ds-\frac az\right) \right) 
$$
$$
=\frac 1{z\sqrt{2\pi t}}\exp \left( -\frac 1{2t}\left( \int\limits_0^t\xi
\left( s\right) \,ds-\frac az\right) ^2\right) \,\vartheta \left( \frac
i{2\pi t}\left( \int\limits_0^t\xi \left( s\right) \,ds-\frac az\right)
\,,\frac i{2\pi tz^2}\right) 
$$
$$
=S\sigma _z\delta \left( \xi \right) \cdot \,\vartheta \left( \frac i{2\pi
t}\left( \int\limits_0^t\xi \left( s\right) \,ds-\frac az\right) \,,\frac
i{2\pi tz^2}\right) \text{ .} 
$$
Thus we have proved:

\ 

{\bf Theorem 3.3.1}:\smallskip\ 

\noindent {\it For all }$a\in {\bf C}$ {\it and all }$z\in {\bf S}_{0\text{ }%
}${\it the infinite sum} $\Phi =\sum\limits_{n=-\infty }^\infty \sigma
_z\,\delta \left( B\left( t\right) -a+n\right) $ {\it exists as a Hida
distribution with} $S$-{\it transform}%
$$
S\Phi \left( \xi \right) =S\sigma _z\,\delta \left( \xi \right) \,\cdot
\vartheta \left( \frac i{2\pi t}\left( \int\limits_0^t\xi \left( s\right)
\,ds-\frac az\right) \,,\frac i{2\pi tz^2}\right) \text{ .} 
$$

{\bf 3.4 Local Time}\bigskip\\\noindent Intuitively the local time should
measure the mean time a Brownian particle spends at a given point $a$.\ In
the White Noise framework such an object can be constructed by simply
integrating Donsker`s delta function. Thus one studies the generalized
process $L\left( \cdot ,a\right) $ given formally by ''Tanaka`s formula'':%
$$
L\left( t,a\right) =\int\limits_0^t\delta \left( B\left( s\right) -a\right)
\,ds 
$$
To show that this integral exists as a Hida distribution we estimate the $S$%
-transform of the integrand for complex parameters $a$:%
$$
\left| S\delta \left( B\left( s\right) -a\right) \left( \xi \right) \right|
\leq \frac 1{\sqrt{2\pi s}}\exp \left( \frac 1{2s}\left| 1_{\left[
0,s\right) }\right| _0^2\left| \xi \right| _0^2+\frac{\left| a\right| }%
ss\left| \xi \right| _\infty -\frac{Re\left( a^2\right) }{2s}\right) 
$$
$$
\leq \frac 1{\sqrt{2\pi s}}e^{\frac 12\left| a\right| ^2}e^{-\frac{Re\left(
a^2\right) }{2s}}\exp \left( \left| \xi \right| _1^2\right) \text{ ,} 
$$
where $\left| \cdot \right| _1$ is defined by (\ref{A}). This fulfills the
conditions of theorem 2.1.3 if $Re\left( a^2\right) >0$. Thus we have an
analytic extension of $L\left( t,a\right) $ to the sector ${\bf S}_0$.

\ \ 

{\bf Theorem 3.4.1}:\smallskip\ \\\noindent $L\left( t,a\right)
=\int\limits_0^t\delta \left( B\left( s\right) -a\right) \,ds$ {\it is a
Hida distribution for} $a\in {\bf S}_0$,{\it \ its }$S$-{\it transform is
given by:}%
$$
SL\left( t,a\right) \left( \xi \right) =\dint\limits_0^t\tfrac 1{\sqrt{2\pi s%
}}\exp \left( -\frac 1{2s}\left( \tint\limits_0^s\xi (\tau )d\tau -a\right)
^2\right) ds\text{ .} 
$$
\ 

\section{\protect\underline{\QTR{bf}{An Application}}\ \protect\smallskip }

\noindent In this section we study a quantum system whose one degree of
freedom is constrained to a circle. Constructing a path integral for such a
system, one has to take into account paths with different winding numbers $n$%
. Thus the following ansatz for the propagator seems to be natural:%
$$
I\left( t,\varphi _1\right) \equiv \sum\limits_{n=-\infty }^\infty
I_0\,\delta \left( \varphi \left( t\right) -\varphi _1+2\pi n\right) \text{ ,%
} 
$$
where $\varphi \left( t\right) =\varphi _0+B\left( t\right) $ is the angle
of position modulo $2\pi $. (Other quantizations would arise if we summed up
the contributions from different winding numbers with a phase factor $%
e^{i\theta n}$ \cite{R}.) However multiplication by $I_0$ corresponds to
complex scaling by $z=\sqrt{i}$ and we have seen in 3.3 that the series does
not converge for this value of $z$. A formal calculation would lead to the
following $S$-transform:%
$$
SI\left( t,\varphi _1\right) \left( \xi \right) =S\sigma _{\sqrt{i}}\delta
_{\left( \varphi _1-\varphi _0\right) }\left( \xi \right) \cdot \vartheta
\left( \frac 1t\left( \sqrt{i}\int\limits_0^t\xi \left( s\right) ds-\left(
\varphi _1-\varphi _0\right) \right) ,\frac{2\pi }t\right) \text{ .} 
$$
However the $\vartheta $-function does not converge for these arguments, see 
\cite{Mu}. To stay within the ordinary White Noise framework we thus
consider as final states smeared wave packets $F$ instead of strictly
localized states. So let 
$$
F\left( \varphi \right) =\sum\limits_{l=-\infty }^\infty a_l\,e^{il\varphi }%
\text{ ,} 
$$
where $\sum\limits_{l=-\infty }^\infty \left| a_l\right| \exp \left( \frac
12s^2l^2\right) <\infty $ for some $s>0$. This leads to:%
\begin{eqnarray*}
I&=&I_0F\left( B\left( t\right) +\varphi _0\right) \\
&=&\sum\limits_{l=-\infty }^\infty a_lI_0^{}\exp \left( il\left( \left\langle
 \omega ,1_{\left[ 0,t\right) }\right\rangle +\varphi _0\right) \right)  .
\end{eqnarray*}It is then easy to calculate%
\begin{eqnarray*}
T\,I\left( \xi \right) & = & \sum\limits_{l=-\infty }^\infty a_{l\,}e^{il\varphi _0}
T\left( I_0\exp \left( il\left\langle \omega ,1_{\left[ 0,t\right) }\right\rangle \right) 
\right) \left( \xi \right) \\
 & = &\sum\limits_{l=-\infty }^\infty a_l\,e^{il\varphi _0}\exp \left( -\frac i2
\int \left( \xi +l1_{\left[ 0,t\right) }\right) ^2d\tau \right) \\
 & = & e^{-\frac i2\int \xi ^2d\tau }\sum\limits_{l=-\infty }^\infty a_{l\,}\exp
 \left( -\frac i2l^2t+il\left( -\int\limits_0^t\xi \left( s\right) ds+\varphi _0\right) \right) 
\text{ .}
\end{eqnarray*}To ensure convergence of the series we estimate:%
\begin{eqnarray*}
  \left| T\,I\left( \xi \right) \right|  & \leq & \sum\limits_{l=-\infty }^\infty \left| a_l
\right| \,e^{ \,\left| l\right| \left| \left( 1_{\left[ 0,t\right) },\xi \right) \right| }
\,e^{\frac 12\left| \xi \right| _0^2}\\
  & \leq & \sum\limits_{l=-\infty }^\infty \left| a_l\right| \,
e^{\left| l\right| \sqrt{t}\left| \xi \right| _0}\,e^{\frac 12\left| \xi \right| _0^2}\\
  & \leq & \left( \sum\limits_{l=-\infty }^\infty \left| a_l\right|
 e^{\frac 12s^2l^2}\right) e^{\frac 12\left( 1+\frac t{s^2}\right) 
\left| \xi \right| _0^2}\text{ .} 
\end{eqnarray*}This is a uniform bound, sufficient for the application of
theorem 2.1.2. Thus we have proved $I\in ${\bf $\left( {\cal S}\right) ^{*}$%
. }It is straightforward to check that the Feynman integral%
\begin{eqnarray*}
 \dint\limits_{S^{\prime }\left( R\right) }I\left( \omega \right) \,d\mu \left( \omega 
\right) & \equiv &\left\langle \left\langle I,1\right\rangle \right\rangle \\
  & = & TI\left( 0\right)  \\
  & = & \dsum\limits_{l=-\infty }^\infty a_l\exp \left( -\frac i2l^2t+il\varphi _0\right) 
\end{eqnarray*}solves the corresponding Schr\"odinger equation.\medskip

{\bf Acknowledgements}

We would like to thank Y. Kondratiev for various helpful discussions. This
work was made possible by financial support from STRIDE.\bigskip\


\begin{thebibliography}{99}
\bibitem{FPS}  De Faria, M. Potthoff, J. Streit, L.: {\it The Feynman
integrand as a Hida distribution. }J. Math. Phys. 32 (1991), 2123-2127.

\bibitem{D}  Doss, H.: {\it Sur une r\'esolution stochastique de
l`\'equation de Schr\"odinger a coefficients analytiques. }Comm. math. Phys.
73 (1980), 247-264.

\bibitem{GV}  Gel'fand, I. M. and Vilenkin, N. Y.: {\it Generalized Functions%
} 4. Academic Press, New York, London, 1964.

\bibitem{H}  Hida, T.: {\it Brownian Motion}. Springer, Berlin, Heidelberg,
New York, 1980.

\bibitem{HKPS}  Hida, T., Kuo, H.-H., Potthoff, J., and Streit, L.: {\it %
White Noise: An Infinite Dimensional Calculus}. Kluwer, Dordrecht 1993.

\bibitem{HKPS2}  Hida, T., Kuo, H.-H., Potthoff, J., and Streit, L. (eds.): 
{\it White Noise - Mathematics and Applications}. World Scientific,
Singapore, 1990.

\bibitem{HS}  Hida, T. and Streit, L.: {\it Generalized Brownian functionals
and the Feynman integral}. Stoch. Processes Appl. 16, (1983) 55.

\bibitem{KS}  Khandekar, D. C. and Streit, L.: {\it Constructing the Feynman
integrand.} Ann. Physik 1 (1992), 49-55.

\bibitem{kon80}  Kondratiev, Yu.G.: {\it Nuclear spaces of entire functions
in problems of infinite dimensional analysis.} Soviet Math. Dokl. 22 (1980),
588-592

\bibitem{KLPSW}  Kondratiev, Y. G., Leukert, P., Potthoff, J., Streit L. and
Westerkamp W.:{\it \ Generalized Functionals in Gaussian Spaces - the
Characterization Theorem Revisited.} (1993), Preprint.

\bibitem{KoS}  Kondratiev, Y. G. and Streit, L.: {\it A remark about a norm
estimate for White Noise distributions}. Ukrainian Math. J. (1992) no.7.

\bibitem{KoS2}  Kondratiev, Y. G. and Streit, L.: {\it Spaces of White Noise
distributions: Constructions, Descriptions, Applications }I. BiBoS Preprint
(1991), to appear in Rep. Math. Phys.

\bibitem{Ku}  Kubo, I.: {\it It\^o formula for generalized Brownian
functionals}. Theory and Application of Random Fields, ed. G. Kallianpur.
Springer, Berlin, Heidelberg, New York (1983).

\bibitem{KT}  Kubo, I. and Takenaka, S.: {\it Calculus on Gaussian White
Noise I+II}. Proc. Japan Acad. 56A (1980), 376-380 and 411-416.

\bibitem{KY}  Kubo, I. and Yokoi, Y.: {\it A remark on the space of testing
random variables in the White Noise calculus}. Nagoya Math. J. 115 (1989),
139-149.

\bibitem{K}  Kuo, H.-H.: {\it Lectures on White Noise Analysis}. Soochow
Univ. Lectures, 1990.

\bibitem{kps}  Kuo, H.-H., Potthoff, J. and Streit, L.: {\it A
characterization of white noise test functionals.} Nagoya Math. J. {\bf 121 }%
(1991), 185-194.

\bibitem{LLSW}  Lascheck, A. Leukert, P. Streit, L. Westerkamp, W.:{\it \
Quantum mechanical propagators in terms of Hida distributions. }Rep. Math.
Phys. {\bf 33} (1993), 221-232.

\bibitem{Lee}  Lee, Y.J.: {\it Generalized Functions of Infinite Dimensional
spaces and its Application to White Noise Calculus}. J. Func. Anal. {\bf 82}%
, (1989), 429-464

\bibitem{MY}  Meyer, P. A. and Yan, J. A.: {\it Les ''fonctions
caract\'eristiques'' des distributions sur l'espace de Wiener}. S\'eminaire
de Probabilit\'e XXV, LNM 1485; Springer, Berlin, Heidelberg, New York.

\bibitem{Mu}  Mumford, D.: {\it Tata Lectures on Theta I}. Birkh\"auser,
Boston, Basel, Stuttgart (1979).

\bibitem{P}  Potthoff, J.: {\it Introduction to White Noise Analysis}. LSU
preprint (1991).

\bibitem{PS}  Potthoff, J. and Streit, L.: {\it A Characterization of Hida
Distributions}. J. Funct. Anal. 101 (1991), 212-229.

\bibitem{PS2}  Potthoff, J. and Streit, L.: {\it Generalized Radon-Nikodym
Derivatives and Cameron-Martin Theory.} In: ''Gaussian Random Fields'',
It\^o, K. and Hida, T. (eds.). World Scientific, Singapore (1991).

\bibitem{R}  Rivers, R.: {\it Path integral methods in quantum field theory.}
Camebridge University Press, Camebridge, New York, Sydney, 1987.

\bibitem{S}  Streit, L.: {\it The Feynman integral - recent results}. In:
''Dynamics of complex and irregular systems'', Eds. Ph. Blanchard et al.,
World Scientific, Singapore, 1993.

\bibitem{SW}  Streit, L. and Westerkamp, W.: {\it A generalization of the
characterization theorem for generalized functionals of White Noise}. In:
''Dynamics of complex and irregular systems'', Eds. Ph. Blanchard et al.,
World Scientific, Singapore, 1993.

\bibitem{Wa1}  Watanabe, H.: {\it The local time of self-intersections of
Brownian motion on generalized Brownian functionals. }Lett. Math. Phys. 23
(1991), 1-9.

\bibitem{Wa2}  Watanabe, H.: {\it Donsker`s delta function and it`s
application in the theory of White Noise Analysis.} Kallianpur Festschrift
Springer (1993), 338.

\bibitem{W}  Westerkamp, W.: {\it A primer in White Noise Analysis}. In:
''Dynamics of complex and irregular systems'', Eds. Ph. Blanchard et al.,
World Scientific, Singapore, 1993.

\bibitem{yan90}  Yan, J.-A., {\it A characterization of white noise
functionals.} Preprint 1990.

\bibitem{Y}  Yosida, K.: {\it Functional Analysis}. Springer, Berlin,
Heidelberg, New York, 1980. \ 
\end{thebibliography}
\end{document}